
%
\input harvmac
\def\S{Schr\"odinger }
\def\pseudoH{\widetilde{\cal H}}
\def\Psiy{\widetilde \Psi}

\def\sqrt#1{(#1)^{1/2}}
\Title{HWS-92/06}{Quantum Mechanics with Explicit Time Dependence}

\centerline{John Rogers\footnote{*}{Address after Sept. 1, 1992:
Department of Applied Physics, Columbia University, New York, NY 10027 USA}
and Donald Spector\footnote{$^\dagger$}
{spector@hws.bitnet}}
\bigskip\centerline{Department of Physics, Eaton Hall}
\centerline{Hobart and William Smith Colleges}
\centerline{Geneva, NY 14456 USA}

\vskip .3in
We investigate quantum mechanical Hamiltonians with explicit
time dependence. We find a class of models in which an analogue
of the time independent \S equation exists. Among
the models in this class is a new exactly soluble model, the
harmonic oscillator with frequency inversely proportional to time.

\Date{July 1992}

In this paper, we study quantum mechanical systems in which the
Hamiltonian has an explicit dependence on time. As is well-known,
when a Hamiltonian is independent of time, the solution of the
corresponding quantum system can be reduced, via separation of
variables, to solving the time-independent \S equation (see for
example \ref\Merzbacher{E. Merzbacher, {\it Quantum Mechanics},
$2^{nd}$ ed. (John Wiley \& Sons, New York, 1970).}).
When the Hamiltonian depends on time, such a simplification does not
generically occur. In this paper, we will
identify a class of time-dependent Hamiltonians for which an analogue
of the time-independent \S equation does exist. We stress that in our work,
the time dependence is treated exactly; we do
not assume the time dependence to be adiabatic or
perturbative. Among the models we consider, we find a new exactly
soluble model: the harmonic oscillator with frequency inversely
proportional to time.

Time-dependent Hamiltonians arise in a variety of situations,
ranging from laboratory setups in which the experimenter dynamically
modifies the environment, to situations in which the strength of an
interaction depends on the expectation value of field which has not
yet reached (or, in the case of a runaway potential, might never reach)
equilibrium. In the context of cosmology, Dirac proposed that the
Newtonian gravitational constant might depend on time, with the implication
that gravity is weak because
the universe is old \ref\Dirac{P.A.M. Dirac, {\it Nature} {\bf 139}
(1937) 323\semi
P.A.M. Dirac, {\it Proc. Roy. Soc.} {\bf A165} (1938) 199.}.
Formally, the problem of determining the excitation spectrum of certain
time-dependent solitons can be formulated as solving particular
quantum systems with time-dependent
Hamiltonians \ref\CL{N.H. Christ and T.D Lee, {\it Phys. Rev.} {\bf D12}
(1975) 1606.}.

Our approach will be first to identify a class of time-dependent
models in which one can obtain an analogue of the
time-independent \S equation, and then to study  one system in detail,
the harmonic oscillator with frequency inversely proportional to time.
For the sake of simplicity, we will restrict our attention in
this paper to systems with only one spatial dimension, but it
will be readily apparent that our techniques can be applied to
higher dimensional models as well.

The \S equation in position space reads
\eqn\tdSe{{\bf H}\Psi(x,t)
   = i\hbar {\partial \Psi(x,t) \over \partial t}~.}
Let us consider the simplest class of time-dependent Hamiltonians,
\eqn\Hgt{{\bf H} = -{\hbar^2 \over 2m}{\partial^2 \over \partial x^2} +
g(t) x^n~.}
As long as $g(t)$ is real, this Hamiltonian is hermitian.
Of course, separation of variables in terms of $x$ and $t$ will not
be fruitful in this case, so we attempt to find a change of
variables for which separation of variables will be useful. To this
end, we define a new variable $y = f(t)x$. The time-dependent
\S equation now reads
\eqn\newSE{-{\hbar^2 \over 2m}f(t)^2 {\partial^2 \over \partial y^2}\Psiy(y,t)
+ {g(t) \over f(t)^n}y^n\Psiy(y,t) = i\hbar{\dot f(t) \over
f(t)}y{\partial \over \partial y}\Psiy(y,t) + i \hbar {\partial
\Psiy(y,t) \over \partial t}~.}
We use $\Psiy$ to denote the wavefunction as a function of $y$ and $t$.
Obviously, $\Psi(x,t)=\Psiy(y(x,t),t)$. Note that in
\newSE, the partial derivatives with respect to time are now taken
with $y$ fixed rather than with $x$ fixed.

Inspecting \newSE, it is easy to see that separation of variables
in $y$ and $t$ will work provided that we choose
\eqn\fcond{f(t)=(pt+q)^{-1/2}~,}
and that
\eqn\gcond{
g(t) = {(pt+q)^{-(2+n)/2} }~,}
where $p$ and $q$ are constants. Hereon, we take \fcond\
and \gcond\ to hold.
By considering solutions to \newSE\
of the form $\Psiy(y)=T(t)Y(y)$, we see that
the general solution to the \S equation can be written as
\eqn\YTlincom{\Psiy(y,t) = \sum_k{c_k T_k(t)Y_k(y)}~,}
where the $c_k$ are arbitrary constants,
and where $Y_k$ and $T_k$
are the solutions of, respectively,
\eqn\Yeqn{-{d^2 \over dy^2}Y_k
 + i{mp \over\hbar}y{d \over dy}Y_k
 + {2m \over \hbar^2}y^n Y_k = \gamma_k Y_k }
and
\eqn\Teqn{ {2mi\over \hbar}(pt+q){d \over dt}T_k
   = \gamma_k T_k~,}
with $\gamma_k$ a constant.

The functions $Y_k$ are formally analogous to the
stationary states of ordinary quantum mechanics, but
are not states of definite energy. The
constants $\gamma_k$ characterize these basic solutions but, again in
contrast to the ordinary case, these constants do {\it not}
correspond to the energies of the associated states. Indeed, the
$\gamma_k$ are complex, with $Im(\gamma_k)$ the same for all $k$.

The differential equation for $T_k$ can
be solved exactly, yielding
\eqn\Tsoln{T_k(t) ={\Bigl(1 + {pt\over q}\Bigr)}^{-(i\hbar\gamma_k/ 2mp)}~,}
where we have chosen the normalization $T_k(0)=1$.
Note that since the $\gamma_k$ are not real, the $T_k$ are not pure phase.

Thus, we have found a class of time-dependent Hamiltonians for
which separation of variables {\it is} useful. For these systems,
one recovers a mathematical structure formally analogous
to the one one finds for time-independent systems. In particular,
if we define the pseudo-Hamiltonian
\eqn\YH{\pseudoH = -{d^2\over dy^2} + i{mp \over\hbar}y{d \over dy}
  + {2m \over \hbar^2}y^n ~, }
we have reduced the problem of solving the time-dependent
\S equation to finding the eigenfunctions and eigenvalues
of $\pseudoH$.
The most significant difference
between the case we consider and the ordinary case of
time-independent potentials is this: in the ordinary case, the operator whose
eigenfunctions and eigenvalues we must find has exactly the same
form as the Hamiltonian; in our case, the operator whose
eigenfunctions and eigenvalues we must find is {\it not} the
Hamiltonian. (Note that in the limit that $p=0$ the
pseudo-Hamiltonian becomes the ordinary Hamiltonian, as it must.)

For the remainder of this paper, we concentrate on a particular
example, the harmonic oscillator with frequency
\eqn\weqn{\omega(t) = {r\over t+s}~,}
where $r$ and $s$ are constants, with $|r|>{1\over 2}$. (We will see the
importance of this restriction later.)
This choice of $\omega(t)$ ensures that the Hamiltonian
\eqn\tdho{ {\bf H} = -{\hbar^2\over 2m}{\partial^2 \over \partial x^2} +
{1\over 2}m\omega^2(t) x^2 }
satisfies the criterion \gcond\ which we derived above. (Note that $s$
reflects the choice for the origin of time, and so, with a suitable
choice, this is a system with frequency inversely proportional to time.)

With the change of variables
\eqn\hoy{y = \Bigl({m\over \hbar(t+s)}\Bigr)^{1/2} x~}
(here we are absorbing extra factors of $\hbar$ and $m$ into $y$,
as compared to \fcond),
separation of variables with $y$ and $t$ will work.
We find that
\eqn\hoT{T_k(t) =\Bigl(1+{t\over s}\Bigr)^{-(i\gamma_k/ 2)}~.}

The analogue of the stationary states for this system satisfy the
ordinary differential equation
\eqn\hoYeqn{
-{d^2 \over dy^2}Y_k +i{d \over dy}Y_k
    + r^2 y^2 Y_k = \gamma_k Y_k~.}
This equation is solved most easily by operator methods.
The pseudo-Hamiltonian operator $\pseudoH$ for this system is
\eqn\fakeham{\pseudoH= -{d^2 \over dy^2}+iy{d \over dy}+r^2 y^2 ~.}
Now define the ladder operators
\eqn\ladder{ a={d\over dy}+\alpha y~, \qquad
   a^\dagger = -{d\over dy}+\alpha^*y ~,}
where $\alpha$ is a complex number. Choosing
\eqn\findalpha{\alpha = r e^{-i\theta}~,
  \qquad \sin\theta = {1\over 2r} ~,}
(recall that $|r|>{1\over 2}$),
we have the algebra \eqna\algebra
$$\eqalignno{
[a,a^\dagger] &= 2r\cos\theta &\algebra a\cr
{1\over 2}\{a,a^\dagger\}&= {\pseudoH} +{i\over 2} &\algebra b \cr
[\pseudoH,a]&= -2r\cos\theta a &\algebra c \cr
[{\pseudoH},a^\dagger]&=2r\cos\theta a^\dagger~.&\algebra d\cr}$$
This algebra is clearly a generalization of the familiar algebra of
the raising and lowering operators of the
time independent harmonic oscillator.

{}From \algebra{b}, we see that $\gamma_k = Re(\gamma_k)-{i\over 2}$,
and that $Re(\gamma_k)\ge 0$. Without loss of generality, we consider the case
the $r\cos\theta > 0$. In this case, we see that $a$ lowers and $a^\dagger$
raises the real part of the $\pseudoH$ eigenvalues by $2r\cos\theta$.
Thus we see that there must be a ``ground state'' $Y_0$ such that
\eqn\ground{a Y_0 = 0~.}
This yields that
\eqn\Yzero{Y_0(y) = \exp(-{1\over 2}\alpha y^2)~,}
while the other states of this system are given (up to
normalization) by
\eqn\allY{Y_k(y) = (a^\dagger)^k Y_0(y)~,\qquad k=1,2,3,\ldots~.}
Note that all our results reduce to the known results for the
time-independent harmonic oscillator
in the appropriate limit ($r \rightarrow \infty$, with $s/r $ finite).

We can evaluate the $\gamma_k$ from \hoYeqn\ using these raising and
lowering operators. One readily finds that
\eqnn\allgamma
$$\eqalignno{\gamma_k
  &= \alpha+ k(\alpha^* + \alpha) \cr
  &= -{i\over 2}+(k+{1\over 2})\sqrt{4r^2-1}~,
    \qquad k=0,1,2,\ldots~.&\allgamma \cr}$$
Thus we have solved completely and exactly the quantum
mechanical harmonic oscillator with frequency inversely proportional to time.
All physically observable quantities may now be computed explicitly
for this system.

In closing, we remark that we have identified time-dependent
quantum systems for which separation of variables is a useful and
effective technique. We have presented this work in one spatial
dimension, but clearly the same methods can be applied in higher
dimensions as well. In the case of the time-dependent harmonic
oscillator, we have identified a new exactly soluble model.
There are several areas appropriate for further development,
including the application of these methods to field theoretic
examples; the identification of solitons whose excitation spectrum
can be found using the techniques we have described; and the
interpretation of our results in a fundamental way in $x$-space.
It would be particularly interesting to classify the exactly
soluble models with the type of time dependence we have explored;
indeed we anticipate
that many familiar exactly soluble time-independent
models have exactly soluble time-dependent generalizations.

\bigbreak\bigskip\bigskip\centerline{{\bf Acknowledgments}}\nobreak
J.R. acknowledges the support of a Hobart and William Smith Summer
Student Research Grant.

\listrefs
\bye